\documentclass[ twocolumn, hf,
]{ceurart}

\usepackage{latexsym}
\usepackage{amssymb}
\usepackage{amsmath}
\usepackage{amsthm}
\usepackage{booktabs}
\usepackage{enumitem}
\usepackage{graphicx}
\usepackage{color}
\usepackage{subfigure}
\usepackage{comment}

\renewcommand{\baselinestretch}{0.985}

\newcommand{\BibTeX}{B\kern-.05em{\sc i\kern-.025em b}\kern-.08em\TeX}

\usepackage{listings}
\conference{ECAI'24: International Workshop on AI in Education and Educational Research (AIEER), October 19--20, 2024, Santiago de Compostela, Spain}

\title{Enhancing Critical Thinking in Education by means of a Socratic Chatbot}

\author[1]{Lucile Favero}[%
orcid=0009-0005-2981-0124,
email=lucile@ellisalicante.org
]
\cormark[1]
\fnmark[1]

\author[2]{Juan Antonio Pérez-Ortiz}[%
orcid=0000-0001-7659-8908,
]
\cormark[2]
\fnmark[2]

\author[3]{Tanja Käser}[%
orcid=0000-0003-0672-0415,
]
\cormark[3]
\fnmark[3]

\author[1]{Nuria Oliver}[%
orcid=0000-0001-5985-691X,
]
\cormark[1]
\fnmark[1]

\address[1]{ELLIS Alicante, Spain}
\address[2]{Universitat d'Alacant, Spain}
\address[3]{École Polytechnique Fédérale de Lausanne, EPFL, Switzerland}

\begin{document}

\maketitle
\begin{abstract}
While large language models (LLMs) are increasingly playing a pivotal role in education by providing instantaneous, adaptive responses, their potential to promote critical thinking remains understudied. In this paper, we fill such a gap and present an innovative educational chatbot designed to foster critical thinking through Socratic questioning. Unlike traditional intelligent tutoring systems, including educational chatbots, that tend to offer direct answers, the proposed Socratic tutor encourages students to explore various perspectives and engage in self-reflection by posing structured, thought-provoking questions. Our Socratic questioning is implemented by fine and prompt-tuning the open-source pretrained LLM with a specialized dataset that stimulates critical thinking and offers multiple viewpoints. In an effort to democratize access and to protect the students' privacy, the proposed tutor is based on small LLMs (Llama2 7B and 13B-parameter models) that are able to run locally on off-the-shelf hardware. We validate our approach in a battery of experiments consisting of interactions between a simulated student and the chatbot to evaluate its effectiveness in enhancing critical thinking skills. Results indicate that the Socratic tutor supports the development of reflection and critical thinking significantly better than standard chatbots. Our approach opens the door for improving educational outcomes by cultivating active learning and encouraging intellectual autonomy.

\end{abstract}


\section{Introduction}

Recent advancements in generative artificial intelligence (AI) methods, combined with an almost universal availability of generative AI tools, have expanded the potential to assist people in developing new ways to think, communicate, learn, and solve problems creatively \cite{henriksen_creative_2022}. The integration of large language models (LLMs) in education through interactive chatbots has emerged as a transformative element in the evolving educational technology landscape, opening new avenues for learning and pedagogy \cite{kuhail2023interacting}. 

While recent work illustrates the potential positive impact of chatbots in learning \cite{ait2023impact}, their capabilities to support essential aspects of human learning, such as critical thinking and self-regulated learning, are limited \cite{lehmann2014cognitive}. However, addressing such limitations is of paramount importance, as these aspects are widely regarded as fundamental pillars in human learning yet inadequately accomplished with current technologies~\cite{liang_let_2023}.

 In this paper, we focus on tackling these limitations. We present a novel Socratic chatbot designed to align with pedagogical principles by promoting critical thinking, purposeful learning and self-efficacy. We propose the use of Socratic dialogues to enhance learning experiences by having the chatbot ask questions rather than provide immediate answers. Our approach combines the traditional elements of Socratic questioning with the innovative capabilities of AI, aiming to create a personalized, responsive and enjoyable learning experience. Furthermore, we are interested in exploring the capabilities of small, open-source LLMs running locally on the learners' laptops to act as Socratic tutors. Locally run LLMs provide more privacy and contribute to the democratization of education as they do not need a constant internet connection benefiting students in rural or underserved areas. 

This paper is organized as follows: the most relevant related literature is presented in Section \ref{sec:litreview}, establishing the background and context for our research. Section \ref{sec:method} describes our approach to develop a Socratic chatbot, consisting of (1) fine-tuning an open-source pretrained large language model on a dataset designed to enhance critical thinking skills and (2) prompt-tuning it to create conversations according to the Socratic method. Our experimental evaluation and comparison of the Socratic tutor with two baselines is described in Section \ref{sec:Evaluation}, followed by the result in Section \ref{sec:Results}, a discussion in Section \ref{sec:discussion} and a conclusion and outline of future lines of research in Section \ref{sec:conclusion}. Furthermore, a discussion about the ethical implications and challenges is provided in the supplementary material.
 
\section{Related work}\label{sec:litreview}
In this section, we first provide an overview of the most relevant previous works in critical thinking and Socratic questioning, both from an educational and technological perspective. 

\subsection{Educational chatbots}
For the past fifteen years, scholars have explored several facets of integrating chatbots into educational settings, examining their effectiveness, benefits, and challenges \cite{perez2020rediscovering}. Early work investigated basic interactions and information retrieval tasks~\cite{kerlyl2006bringing}. As AI methods improved, studies explored the role of chatbots in personalized learning \cite{smutny2020chatbots}, student engagement \cite{goda2014conversation}, and language learning \cite{fryer2017stimulating}. Recently, the use of LLM-based chatbots in education has gained significant research attention and societal adoption with the emergence of numerous EdTech startups devoted to this topic \cite{rudolph2023war}. Yet, there is a lack of educational chatbots to support critical thinking and self-regulated learning. 

\subsection{Critical thinking}
\paragraph{Definition in education.}
Critical thinking (CT) is a complex cognitive process involving the evaluation, analysis, and synthesis of information to form reasoned judgments \cite{abrami2015strategies, alsaleh2020teaching, paul_thinkers_2019}. Essential across all educational levels and disciplines, CT empowers students to think independently, solve problems, and make informed decisions. Several strategies are effective in promoting CT skills \cite{abrami2015strategies}, such as: (1) engaging students in structured dialogues to explore various perspectives; (2) exposing students to real-world scenarios requires them to actively apply their CT skills;  and (3) providing personalized mentoring to support the development of CT capabilities.

However, assessing CT skills presents difficulties due to a lack of a standardized evaluation framework, complicating efforts to further integrate and enhance CT training in educational environments \cite{alsaleh2020teaching}.

\paragraph{Chatbots for critical thinking.}
Incorporating chatbot technology into educational settings offers substantial benefits for enhancing CT \cite{parsakia2023effect}. Indeed, chatbots can effectively employ the three previously mentioned strategies by facilitating guided dialogues focused on specific, concrete problems. Although research in this area is still emerging, some progress has been reported in the past two years.  
For example, the AI-IESLS chatbot enhances the students' learning experiences and critical thinking in higher education through the use of concept mapping and probability distribution analysis \cite{Liu2022AI}. Moreover, a chatbot designed according to cognitive styles has been proven effective in advancing science concepts and critical thinking skills among preparatory school students \cite{Barsoum2022The}.Finally, the online education nonprofit Khan Academy has recently developed Khanmigo, an AI-powered tutor and teaching assistant hand-crafted by their learning experts. It aims to promote critical thinking by guiding students through interactive problem-solving.\footnote{https://www.khanmigo.ai}

\subsection{Socratic questioning}

\begin{table}[ht]     
\centering    
\caption{Description and examples of each Socratic Question-Type reproduced from  \cite{ang_socratic_2023}. *Other refers to questions that do not conform to Socratic categories.}
\begin{tabular}{p{1.5cm}p{3cm}p{1.5cm}}
\toprule
\multicolumn{1}{c}{\textbf{Socratic}} & \multicolumn{1}{c}{\textbf{Description}}& \multicolumn{1}{c}{\textbf{Example}}\\ 
\multicolumn{1}{c}{\textbf{question type}} & & \\ 
\hline \\ 
Clarification  & Question probing the ambiguities of a thought. & What do you mean by ...?\\[0.3ex] \hline \\[0.1ex] 
Probing assumptions  & Question probing the assumptions behind a thought. &Why do you assume ...?\\[0.3ex] \hline \\[0.1ex] 
Probing reasons and evidences  & Question probing the justifications or concrete evidences that have supported a thought. &How did you know that ... ? \\[0.3ex]\hline \\[0.1ex] 
Probing implications and consequences  & Question probing the impacts or implications of a thought. &If ..., what is likely to happen as a result?\\[0.3ex]\hline \\[0.1ex] 
Probing alternative viewpoints and perspectives  & Question probing other possible viewpoints. &What else should we consider about...?\\[0.3ex]\hline \\[0.1ex] 
Other*  & Question unrelated to the question types above (e.g. rhetorical, irrelevant, and/or illogical questions, etc.). &Who wouldn’t want to be rich?\\[0.1ex]

\bottomrule
\end{tabular}
\label{tab:Questions_type}    
\end{table}

\paragraph{Definition in education.}
Named after the classical Greek philosopher Socrates, Socratic questioning is a structured way of asking questions to explore complex ideas, reveal underlying assumptions, and discern what is known from what is not \cite{paul_thinkers_2019}.  

Socratic questioning is part of the scaffolding theory of learning, a process in which a learner achieves a goal through guided efforts~\cite{wood1976role}. Students are prompted with questions that guide them towards the solution by solving problems beyond their zone of proximal development \cite{vygotsky2012thought}.

Its primary purpose in education is to test and deepen the students' understanding, encourage them to ask profound questions, help them think critically about complex topics, and foster an understanding of different viewpoints \cite{paul_thinkers_2019}.
Socratic questioning has been shown to promote critical thinking, improve the student's understanding and encourage a culture of inquiry and reflection, essential for reasoning through complex issues and understanding diverse perspectives \cite{zare_use_2015, ho2023thinking, dalim2022promoting}. 

However, the effectiveness of the Socratic method varies among learners because it depends on their ability to abstract and generalize knowledge~\cite{goldin_producing_2017}. 
Thus, this method should be adapted to individual differences in cognitive processing, prior knowledge, or learning styles \cite{paul_thinkers_2019}.

\paragraph{Chatbots for Socratic questioning.}
Chatbots can support the Socratic method of questioning by engaging users in a guided dialogue, consisting of probing, open-ended questions that encourage critical thinking and deep reflection \cite{shridhar_automatic_2022, alsaleh2020teaching}. 

To the best of our knowledge, we are unaware of any fully implemented chatbot designed to leverage the Socratic questioning method to enhance critical thinking in human learning. Nonetheless, recent advancements have applied machine learning techniques (like zero-shot prompt tuning and reinforcement learning) to automatically generate relevant Socratic questions in diverse contexts: math word problems \cite{shridhar_automatic_2022}, debugging code \cite{al2024can}, physics~\cite{gregorcicchatgpt}, and cognitive behavioral therapy~\cite{izumi2024response}. 

Furthermore, Ang et al. \cite{ang_socratic_2023} introduced \textit{SocratiQ}, a dataset consisting of approximately $110,000$ Socratic question-context pairs curated from the subreddit \textit{r/changemyview}.\footnote{\url{https://www.reddit.com/r/changemyview/}} The authors employed a transformer-based encoder to classify each question-context pair into specific Socratic question types as detailed in Table \ref{tab:Questions_type}. Subsequently, for each identified type of Socratic question, a decoder-like language model was trained to generate new questions corresponding to that specific type from a given context. Both automated and human evaluation studies demonstrated that these models successfully produce type-sensitive, relevant, and human-like Socratic questions.

Filling the existing gap in the literature, our Socratic tutor's core design principle is the use of Socratic questioning to guide the student's learning process, changing the focus from providing answers to formulating questions that would support the learner's critical thinking. In our approach, we leverage the \textit{SocratiQ} annotated dataset of question-context pairs to fine-tune an open-source, pretrained LLM to generate specific types of Socratic questions.

\subsection{Contributions}
Given the existing literature, the contributions of our paper are five-fold:
\begin{enumerate}
\item We develop an LLM model focused on critical thinking by fine and prompt-tuning open-source, pretrained small LLMs (Llama2 7B- and 13B-parameter models) to create a Socratic tutor that poses a variety of Socratic questions in a computationally efficient manner;
\item We propose an approach for generating tutoring dialogues by means of an LLM-based dual-agent (tutor-learner) setup;
\item We introduce an extrinsic evaluation method that assesses the quality of the questions by comparing the learner’s responses to the ideal responses based on the problem description; 
\item We implement an automatic evaluation mechanism using an LLM that is designed to replicate human evaluative behaviors; and 
\item We validate our approach on a Theory of Knowledge task and compare it with two baselines. 
\end{enumerate}

\section{Socratic tutor implementation}\label{sec:method}

In this section, we describe the fine and prompt-tuning approach that we adopted to build our Socratic tutor. We experiment with two open-source, pretrained large language models:  Llama2 Instruct with 7B parameters and Llama2 Instruct with 13B parameters \cite{touvron2023llama} which are fine-tuned on Google Colab with a L4 GPU. The fine-tuned models are then run on an Apple M1 Pro laptop with 32GB RAM. This approach of using small models on a local computer is chosen to enable students to operate the chatbot independently on their personal computers, with the aim of democratizing education and protecting privacy. A key challenge in this regard lies in optimizing the chatbot's performance to ensure it operates efficiently on local machines without excessive resource consumption.

\subsection{Socratic fine-tuning}

To fine-tune the open-source, pretrained LLMs we applied parameter-efficient fine-tuning (PEFT) techniques: Low-Rank Adaptation (LoRA) \cite{hu2021lora} and Quantized Low-Rank Adaptation (QLoRA) \cite{dettmers2024qlora}. They offer a way to effectively adapt LLMs to specific tasks without the need to retrain or alter the entire pre-trained model. LoRA focuses on modifying a small number of additional parameters that act upon the pre-trained weights of the model. This approach is especially valuable due to its efficiency and the reduced computational overhead it introduces. Building upon LoRA, QLoRA incorporates quantization into the fine-tuning process (4-bit quantization in our case). It reduces the precision of the low-rank parameters to further decrease memory usage and computational demands.

\paragraph{Hyperparameters.}

LLMs were fine-tuned on a single GPU, using mini-batches of size 4. We adopted a learning rate of $2 \times 10^{-4}$ and used the AdamW optimizer ($\beta_1 = 0.9$, $\beta_2 = 0.999$) in 32-bit precision and a cosine-based scheduler with a step ratio for linear warmup of $0.03$. A weight decay of $0.001$ was applied to all weights except biases and normalization layer parameters. Regarding QLoRA, the rank of LoRA modules, $r$ is $64$, the LoRA scaling factor, $\alpha$ is $16$ with $0.1$ dropout and 4-bit quantization.

\paragraph{Fine-tuning dataset.}
The fine-tuning process requires a specialized dataset. In our case, we used the \textit{SocratiQ} dataset \cite{ang_socratic_2023} which collects contentious viewpoints and the questions challenging these viewpoints from Reddit (Subreddit  
channel called \textit{subreddit r/changemyview}). This channel corresponds to an active community specifically designed as ``\textit{a place to post an opinion you acknowledge may be flawed, in an effort to understand other perspectives on the issue.}'' Discussions often revolve around controversial topics, such as politics, media, and culture. Members are encouraged to approach these discussions with an openness to dialogue rather than debate. Due to its foundational goal of encouraging users to ``change their view'', comments frequently include questions that promote introspection and reflection.

Ang et al.~\cite{ang_socratic_2023} classified the context-question pairs of this Subreddit into the Socratic categories devised by Paul and Elder \cite{paul_thinkers_2019}. The question types along with definitions and examples can be found in Table \ref{tab:Questions_type}. These categories address intellectual standards to enhance the rigor of Socratic dialogues \cite{paul_thinkers_2019}. 

The fine-tuning was performed with a subset of $600$ context-question types of questions from \textit{SocratiQ} that had been annotated through the Amazon Mechanical Turk crowdsourcing platform~\cite{crowston2012amazon}. Given a context and a specific type of question, the LLMs are fine-tuned to generate Socratic questions of this kind.

\subsection{Socratic prompt-tuning}

In addition, we prompt-tuned the fine-tuned LLMs. Prompt-tuning is a method of fine-tuning a language model on specific tasks by adjusting its input prompts without altering the underlying model. 
We prompt-tuned three different types of LLM-based tutors: the Socratic tutor and two baselines (described in Section \ref{sec:baselines}) which correspond to a basic tutor, and a random tutor. The prompts provided to the Socratic tutor are composed of the items listed below\footnote{We direct the reader to the supplementary material for a description of the prompt engineering performed in this paper.} and reflected in the first row of Table~\ref{tab:Tutor_prompt}:

\begin{itemize}
    \item \textbf{Role}. In the first part of the prompt, we inform the LLM of its role. In the case of the Socratic tutor, it is instructed to be a philosopher who follows Socratic methods, which involves asking probing questions to encourage deep thinking and challenge assumptions. 
    \item \textbf{Goal}. The second part contains the goal of the interaction, namely to help a student think about a question (\textit{Theory of Knowledge or ToK question}).
    \item \textbf{Resources}. Then, the prompt imposes the restriction of using only the provided dialogue history (\textit{history}) and the student's latest response (\textit{input}). These elements serve as the context and content from which the LLM must derive its question, underscoring that the tutor should not teach or introduce new material, but instead focus on exploring the students' thoughts and facilitating their reflection, consistent with the principles of Socratic dialectic \cite{paul_thinkers_2019}. 
    \item \textbf{Expected outcome}. Finally, the prompt specifies the expected outcome. The question asked should logically continue the ongoing conversation and must be designed to lead the student to explore their thoughts and ideas further rather than providing direct answers.
\end{itemize}

Table \ref{tab:Socratic_tutoring} provides an example of a simulated conversation with our \textit{Socratic tutor}.

\begin{table}[ht]          
\centering
\caption{ Example of a Socratic conversation with the proposed Socratic tutor.}
\begin{tabular}{|p{0.6cm}p{5.5cm}|}
 \hline &\\[0.6ex] 
Learner& Help me think about the question: Is replicability necessary in the
production of knowledge?\\[2ex] 

Tutor&  What is it that you are doing when you are producing new information, and
how do you know if you are producing something new, or simply reproducing
something old?\\[1ex] 

 Learner&  When we produce new information, we are combining existing
knowledge and ideas in novel ways to create something original and new. We know
we are producing something new when it differs from what already exists, such as
a unique solution to a problem or a fresh perspective on an old topic.\\[1ex] 

Tutor&  I believe you are asking about whether replicability is necessary in the
production of knowledge, not just whether it can be helpful (though I do think
there are cases where replication isn't possible).\\[1ex] 

 Learner&  Replicability is necessary in producing knowledge.\\[1ex] 

Tutor&  You are assuming a lot of things, so I'll just point them out. Firstly, how
do you know it is something new when you produce it? Is there any way to prove
it is novel or did you simply decide it was new because it felt like it?
Secondly, can we even be sure that humans are capable of producing novel
knowledge in the first place?\\[1ex] 

 Learner&  You're right, I'm making assumptions. How do we know something is
new without proof? We can't be sure if humans can produce novel knowledge.\\[1ex] 

Tutor&  Why do you believe humans are capable of producing novel knowledge in the
first place? And what constitutes proof that something is new or not new? It
seems like a circular argument to say "novelty" when it's just a subjective
feeling that something's different from everything else.\\[1ex] 

 Learner&  Humans are capable of producing novel knowledge because we have
unique cognitive abilities, like creativity and problem-solving skills, that
allow us to generate new ideas and concepts. Proof of novelty lies in the
ability to demonstrate how something is different from existing knowledge,
through objective measures such as experimental data or logical reasoning.\\[0.6ex] 
\hline 
\end{tabular} 
\\[0.6ex] 
\label{tab:Socratic_tutoring}    
\end{table}

\section{Automatic evaluation}\label{sec:Evaluation}
To evaluate the effectiveness of the proposed Socratic tutor in fostering critical thinking, we simulate a dialogue between our Socratic tutor and an LLM-based simulated learner whose task is to respond to a specific Theory of Knowledge (ToK, see below) question. While an empirical evaluation with human subjects would be ideal and it is planned as a next step, simulated evaluations are necessary for several reasons. First, they enable a preliminary assessment of the functionality, efficiency, and safety of the approach. Second, they allow for the creation of benchmark data for comparison against human interaction. Finally, they enable running numerous simulated conversations to gather data and refine the chatbot's behavior without the logistical challenges associated with human participants.

\subsection{Domain: Theory of Knowledge (ToK)}

The \textit{Theory of Knowledge} (ToK) course is part of the International Baccalaureate (IB) Diploma Programme.\footnote{\url{https://www.ibo.org/}} It aims to develop critical thinking by exploring the process of knowing rather than teaching specific knowledge.\footnote{\url{https://www.ibo.org/programmes/diploma-programme/curriculum/dp-core/theory-of-knowledge/what-is-tok/}} The course's main goals include linking knowledge construction to academic disciplines and the real world, fostering awareness of diverse cultural perspectives, and encouraging personal reflection on beliefs and assumptions. 

However, ToK courses and assessments, do not fully capture the depth of critical thinking skills intended, raising concerns about their validity \cite{hughes2014theory}. 
To enhance its effectiveness, the course could benefit from a broader inclusion of critical thinking skills, particularly in creative and active thinking, and adjustments to its assessment strategies to better align with these skills \cite{hughes2014theory}. A chatbot designed to promote critical thinking could engage and support students in diverse and complex questioning, encouraging them to explore and evaluate multiple perspectives and scenarios actively.
Specifically, the final assessment for this course comprises two main components: an exhibition and an essay. The latter, limited to $1,600$ words, addresses a theoretical question within ToK. It typically prompts students to analyze and discuss claims about the nature of knowledge production. An example essay topic might require students to explore the assertion that the methods employed in generating knowledge are influenced by the intended application of that knowledge. \footnote{\url{https://www.ibo.org/programmes/diploma-programme/curriculum/dp-core/theory-of-knowledge/what-is-tok/}} Table~\ref{tab:ToK_eg_5} includes examples of ToK questions from the International Baccalaureate curriculum. \footnote{\url{https://ibtokessaytutor.com/ibtokessaytopics/May2023/ib-tok-essay-titles-May-2023.shtml}}

Our approach is to use our Socratic chatbot to help a learner think about these ToK questions to help them develop and articulate logically their ideas to complete the essay.  

\begin{table}[ht]  
\centering  
\caption{ Examples of ToK questions from the IB curriculum.} 
\begin{tabular}{|p{0.05cm}p{6.5cm}|}
\hline &\\[0.3ex] 
1&Is replicability necessary in the production of knowledge?\\[0.7ex]
2&For artists and natural scientists, which is more important: what can be explained or what cannot be explained?\\[0.7ex]
3&Does it matter if our acquisition of knowledge happens in 'bubbles' where some information and voices are excluded?\\[0.7ex]
4&Do you agree that it is astonishing that so little knowledge can give us so much power?\\[0.7ex]
5&Are visual representations always helpful in the communication of knowledge?\\[0.7ex] 
\hline
\end{tabular}
\label{tab:ToK_eg_5} 
\end{table}

\subsection{Generation of synthetic conversations using LLMs}
In order to empirically evaluate the proposed Socratic tutor we generate synthetic dialogues using two LLM-based agents that engage in a conversation with each other, as has been recently done in the literature with the purpose of assisting users in building linear programming models \cite{abdullin2024synthetic}.
In our case, we create an LLM-based conversational student that generates the learner's responses by prompt-tuning a small LLM (Llama2 7B-parameter model) with the following prompt: ``ANSWER to the question <input> IN ONE SENTENCE. DO NOT USE COMPLEX WORDS OR IDEAS''.

\subsection{Baselines: non-Socratic tutors}\label{sec:baselines}
To evaluate the effectiveness of the proposed Socratic tutor, we benchmark it against two non-Socratic LLM-based tutoring methods: (1) A \textit{basic tutor} which engages learners in discussions about the Theory of Knowledge questions without a specific intention to help them; and (2) a \textit{random tutor} which produces arbitrary responses and serves as a control in our study. 
We create these two tutors by prompt-tuning the open-source, pretrained Llama2 7B-parameter model according to the prompts described in Table \ref{tab:Tutor_prompt}. As a result, we create three different tutors (Socratic, basic and random). 

Additional results comparing the Socratic tutor and a baseline tutor that had not been neither fine- nor prompt-tuned can be found in the supplementary material.

  \begin{table}[ht]     
\centering
\caption{Tutor prompt-tuning of three LLM-based tutors used in this paper: Socratic, basic, and random tutors.}  

\begin{tabular}{p{0.6cm}p{5.7cm}}
\toprule
\multicolumn{1}{c}{\textbf{Tutor}} & \multicolumn{1}{c}{\textbf{Prompt}}\\[0.5ex] \hline \\[0.1ex] 
Socratic tutor & You are a strict Socratic philosopher. 
                Your goal is to help the student think about the question:
                <ToK question>
                Use ONLY information of the dialogue history: <history>
                and the new student response: <input> to 
                ASK THE STUDENT ONE SHORT Socratic question that triggers the reflection
                and logically follow the conversation.\\[1ex] 
\hline \\[0.5ex] 
Basic tutor &   Dialogue history: <history>
                New human speech: <input>
                ANSWER TO THE HUMAN BUT DO NOT HELP HIM. ONE SHORT SENTENCE. DO NOT REPEAT YOURSELF \\[1ex] 
Random tutor &  You are having a conversation with a human. New human response: <input><history>
                Just generate a VERY SHORT random and meaningless text\\
                
\bottomrule
\end{tabular}
\label{tab:Tutor_prompt}    
\end{table}

\begin{table}[ht]          
\centering 
\caption{ Example of a conversation with the basic tutor.} 
\begin{tabular}{|p{0.6cm}p{5.5cm}|}
\hline &
\\ [0.6ex] 
Learner& Help me think about the question: Is replicability necessary in the
production of knowledge?\\[1ex] 

Tutor&  Hmm... Interesting question. Replicability is crucial in the production of
knowledge, as it allows for the verification and validation of research findings
through independent reproduction by other scientists.\\[1ex] 

 Learner&  Replicability is important because it lets other scientists check
if a study's results are real or not.\\[1ex] 

Tutor& Exactly! It's crucial for building a solid foundation of knowledge and
ensuring that research results are reliable and accurate.\\[1ex] 

 Learner&  Replicability is important for verifying the authenticity of
research findings.\\[1ex] 

Tutor&  Indeed, replicability is essential for ensuring the validity and
reliability of research findings.\\[1ex] 

Learner& Replicability is important for verifying research results are
accurate and trustworthy.\\[1ex] 
\hline
\end{tabular}
\label{tab:No_Socratic_tutoring}    
\end{table}

\begin{table}[ht]          
\centering
\caption{ Example of a conversation with the random tutor.}
\begin{tabular}{|p{0.6cm}p{5.5cm}|}
\hline &\\[0.6ex]
Learner& Help me think about the question: Is replicability necessary in the
production of knowledge?\\[1ex] 
Tutor&  Gelatinous hamsters on hoverbikes: A future possibility?\\[1ex] 
Learner&  Gelatinous hamsters on hoverbikes? Hmm, probably not a good idea..
they might get stuck in the hoverbike's fan blades!\\[1ex] 
Tutor&  Fluffy unicorns on roller skates: A new mode of transportation?\\[1ex] 
Learner&  Hell yeah, that's a thing now!\\[1ex]
\hline
\end{tabular}
\\[1.2ex]
\label{tab:Random_talk}    
\end{table}

We test each of these tutors on 5 different ToK questions displayed in Table~\ref{tab:ToK_eg_5}. For each tutor and question, we perform 20 conversations of 5 turns each. Given that the topic of the  conversations is cognitively demanding and thought-provoking, our Socratic chatbot should enable students to successfully respond to the task within a limited number of interactions. Thus, we set the number of turns to 5. We also include results corresponding to longer conversations with 20 turns in Fig.~\ref{fig:Longer_conv} in the supplementary material.

\subsection{Evaluation metrics}
We compare the performance of the three types of chatbots by means of an extrinsic evaluation approach 
\cite{abdullin2024synthetic}.
We assess the quality of the questions generated by the tutors by analyzing the responses provided by the simulated learners. This evaluation operates under a key hypothesis: if the learner delivers accurate responses, this suggests that the tutor has posed effective questions that led the learner to generate such answers. Thus, we measure the effectiveness of the questions by comparing the concatenated learner's responses to a summary of the essential points that are expected in the responses to each of the 5 ToK questions. Table~\ref{tab:ToK_eg}  (as well as Tables \ref{tab:ToK_eg_other} and \ref{tab:ToK_eg_other_continued} in the supplementary material) include examples of the ToK questions used with a summary of the information that should be included in the answer to the questions.

\begin{table}[ht]  
\centering
\caption{ Example of a Theory of Knowledge question with a summary of the information that should be included in the answer.} 
\begin{tabular}{p{7cm}}
\toprule
\textbf{Is replicability necessary in the production of knowledge?}\\[1ex] 
\hline
\\ [0.4ex] 
Replicability is crucial in the production of knowledge as it allows for the
verification and validation of research findings by other scientists. The
distinction between necessary and sufficient requirements highlights the
importance of replicability in scientific research. Replicability ensures that
findings are reliable, trustworthy, and can be verified by other researchers.
The relation between replicability and objectivity is important to consider as
it ensures that research findings are not influenced by personal biases or
preconceptions. Replicability enables research findings to be shared and
disseminated more widely, leading to a more comprehensive understanding of the
topic under study. Prioritizing replicability maintains scientific research's
objectivity, reliability, and credibility.\\[1ex] 
\bottomrule
\end{tabular} 
\label{tab:ToK_eg}    
\end{table}

 We use standard n-gram-based metrics ---BLEU \cite{Papineni02bleu:a,lin-och-2004-orange}, ROUGE-L \cite{lin-2004-rouge} and METEOR \cite{banarjee2005}--- to evaluate the quality of the responses provided by the learners \cite{pan2019recent}. However these commonly-used measures struggle to accurately assess paraphrases that, while different in wording, are equally valid interpretations of the reference questions. Thus, we use the BERTScore \cite{bert-score} to overcome these limitations. Metrics leveraging BERT-based models have demonstrated robustness in evaluating text generation tasks, including summarization and translation, by providing more nuanced assessments of textual similarity \cite{bert-score,he2021deberta}. Finally, we propose a novel evaluation metric by means of an LLM \cite{abdullin2024synthetic} to specifically assess the level of critical thinking in the learner's response. The highest possible value of these metrics is 1. Below is a summary of the advantages and limitations of the BERTScore and LLM-based metrics. See the supplementary material for the other metrics.

\begin{itemize}
    
    \item \textbf{BERTScore:} Originally, the BERTScore \cite{bert-score} leveraged pre-trained contextual embeddings from BERT to evaluate the similarity between words in candidate and reference sentences using cosine similarity. This metric effectively aligns with human judgment, demonstrating strong correlation in both sentence-level and system-level evaluations \cite{bert-score}.
    
    However, \textit{DeBERTa} \cite{he2021deberta} enhances both the BERT and RoBERTa architectures by incorporating disentangled attention mechanisms and an improved mask decoder. With 80 GB of training data, DeBERTa surpasses both BERT and RoBERTa on the majority of natural language understanding tasks, thereby demonstrating superior performance across a broad range of evaluations.

    \item \textbf{LLM-score:} Previous metrics primarily assessed the similarity between the learner's response and a summary of potential answers to the formulated question. To more directly evaluate critical thinking, we introduce a score generated by a pretrained Llama2 7B-parameter LLM~\cite{touvron2023llama} which is specifically prompted to assess the level of critical thinking in the learner's response. The evaluation prompt is composed of the following elements: (1) \emph{role}, provided in the first part of the prompt where it specifies that the LLM should act as an AI evaluator specialized in assessing critical thinking in discourse; (2) \emph{definition of critical thinking} as per \cite{black2008critical}; (3) \emph{goal}, which is expressed as the evaluation of the use of critical thinking in the learner's response to a specific ToK question; and (4) \emph{outcome}, which is a score ranging from 0 to 5, reflecting the degree of critical thinking evident in the responses. This score should be accompanied by an explanation, employing the chain-of-thought prompting technique. Ultimately, the LLM-score is normalized to a scale between 0 and 1 to be comparable to the rest of scores. Table \ref{tab:LLM_Scoring_prompt} provides a summary of the prompt used to generate the LLM-score.
\end{itemize}

\begin{table}[ht]  

\centering  
\caption{Prompt to generate the LLM-score using a pretrained Llama2 7B-parameter model.} 
\begin{tabular}{|p{7cm}|}
\hline
\\ [0.3ex] 
You are an AI evaluator specializing in assessing if a discourse uses critical thinking.\\
Here a definition of critical thinking:
It is the analytical thinking which underlies all rational discourse and enquiry. It is characterised by a meticulous and rigorous approach. As an academic discipline, it is unique in that it explicitly focuses on the processes involved in being rational. These processes include: analysing arguments, judging the relevance and significance of information, evaluating claims, inferences, arguments and explanations, constructing clear and coherent arguments, forming well-reasoned judgements and decisions.\\
Carefully check if the answers of the question: <ToK question>  use critical thinking. Your primary goal is to rate the answers based on the use of critical thinking. The provided answers <Learner response>.\\
PROVIDE THE ANSWER IN A JSON FORMAT WITH FOLLOWING FIELD:
"Score" -int | Score from 0 to 5 Then explain your rating. \\[2ex]
\hline
\end{tabular}
\label{tab:LLM_Scoring_prompt}  
\end{table}

The code to create the LLM-based tutors and learner, and to run the experiments described in this paper can be found here: \url{https://github.com/lucilefavero/ECAI_2024}

\section{Results} \label{sec:Results}
Tables \ref{tab:Socratic_tutoring}, \ref{tab:No_Socratic_tutoring} and \ref{tab:Random_talk} provide examples of simulated conversations with the \textit{Socratic tutor}, the \textit{basic tutor}, and the \textit{random tutor}, respectively, all generated with the Llama2 7B-parameter LLM. We also create another Socratic tutor using the Llama2 13B-parameter LLM to shed light on the impact of increasing the complexity of the model to achieve the task.
The performance of these tutors ---quantified using the BLEU, ROUGE-L, METEOR, BERTScore and LLM score-- are presented in Table \ref{tab:res}, corresponding to the average values of 20 conversations of 5 turns for each 5 of the ToK questions. The differences in performance between the Socratic tutors (indicated with bold font) and the non-Socratic tutors are statistically significant (t-test with p-value < 0.001).

\begin{figure}[ht]
    \centering
    \includegraphics[width=1.1\linewidth]{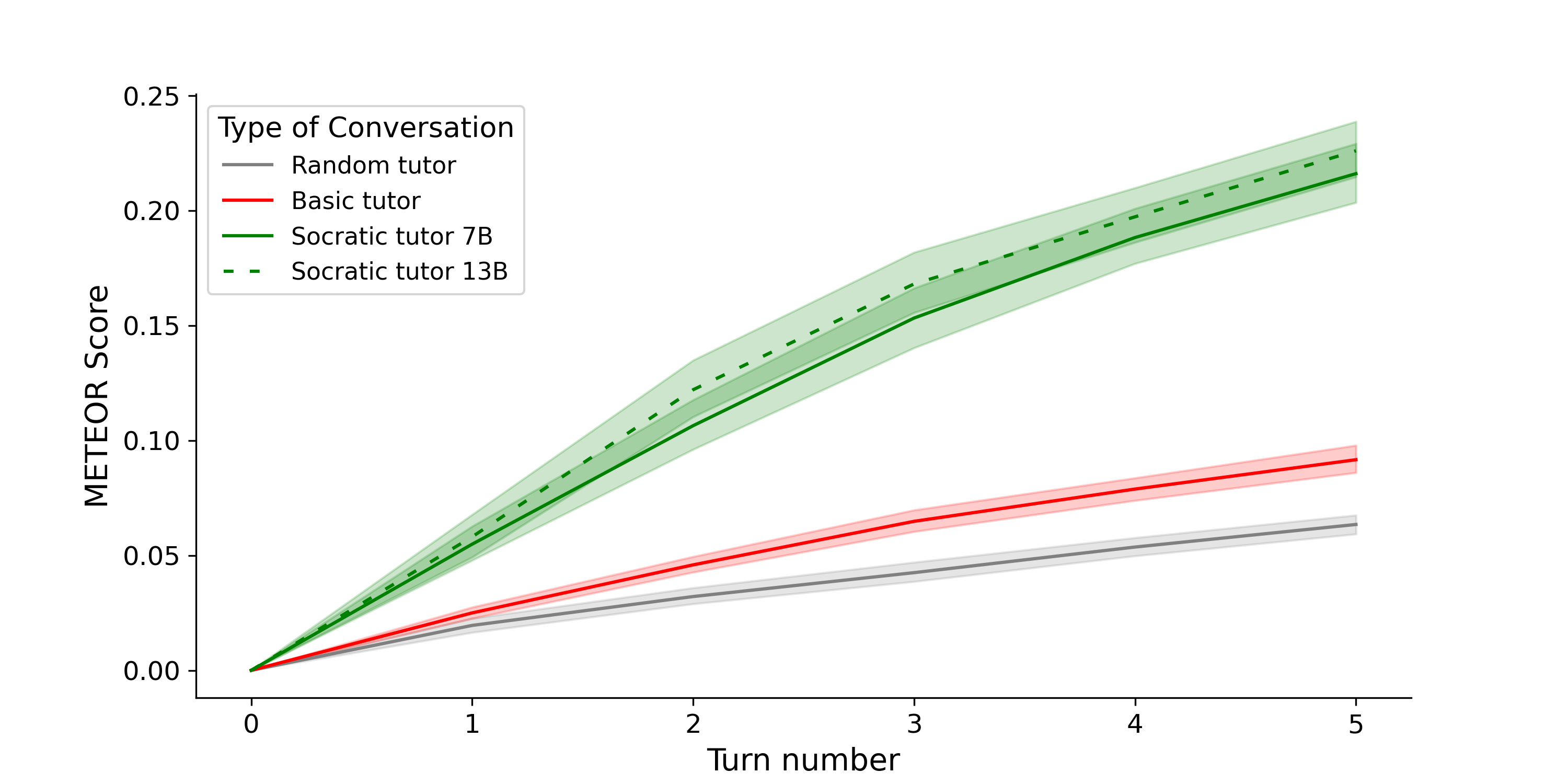}  
    \caption{METEOR scores of the different types of tutors, averaged over 20 conversations of 5 turns each and for each of the 5 ToK questions. The differences between the Socratic tutor and the non-Socratic tutors are statistically significant (t-test, p-value<0.001). No statistically significant difference is observed in the performance of both Socratic tutors.}
    \label{fig:meteor}
\end{figure}

\begin{figure}[ht]
    \centering
    \includegraphics[width=1.1\linewidth]{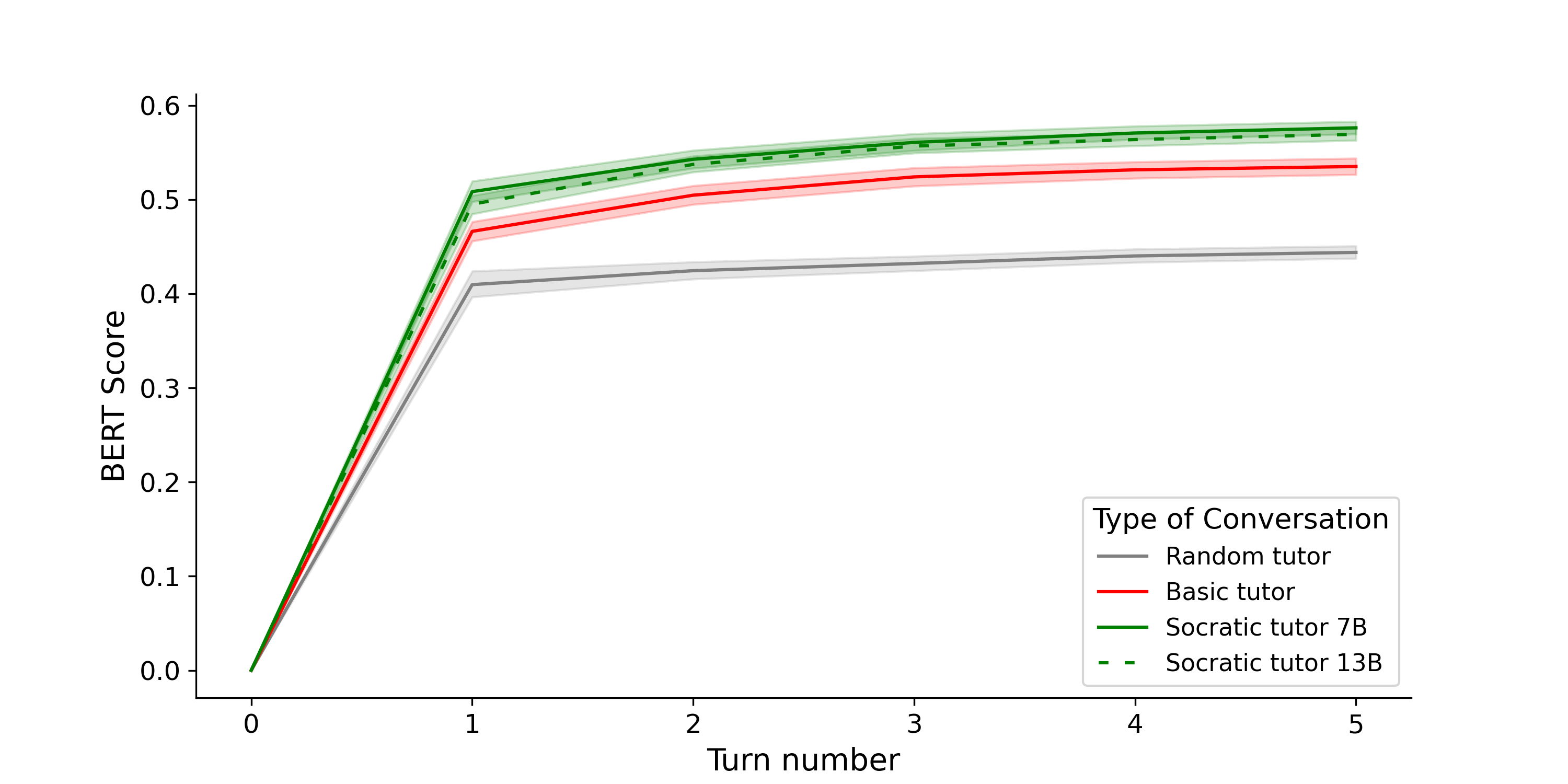}  
    \caption{BERT scores of the different types of tutors, averaged over 20 conversations of 5 turns each and for each of the 5 ToK questions. The differences between the Socratic tutors and the non-Socratic tutors are statistically significant (t-test, p-value<0.001). No statistically significant difference is observed in the performance of both Socratic tutors.}
    \label{fig:bertscore}
\end{figure}

\begin{figure}[ht] 
    \centering
    \includegraphics[width=1.1\linewidth]{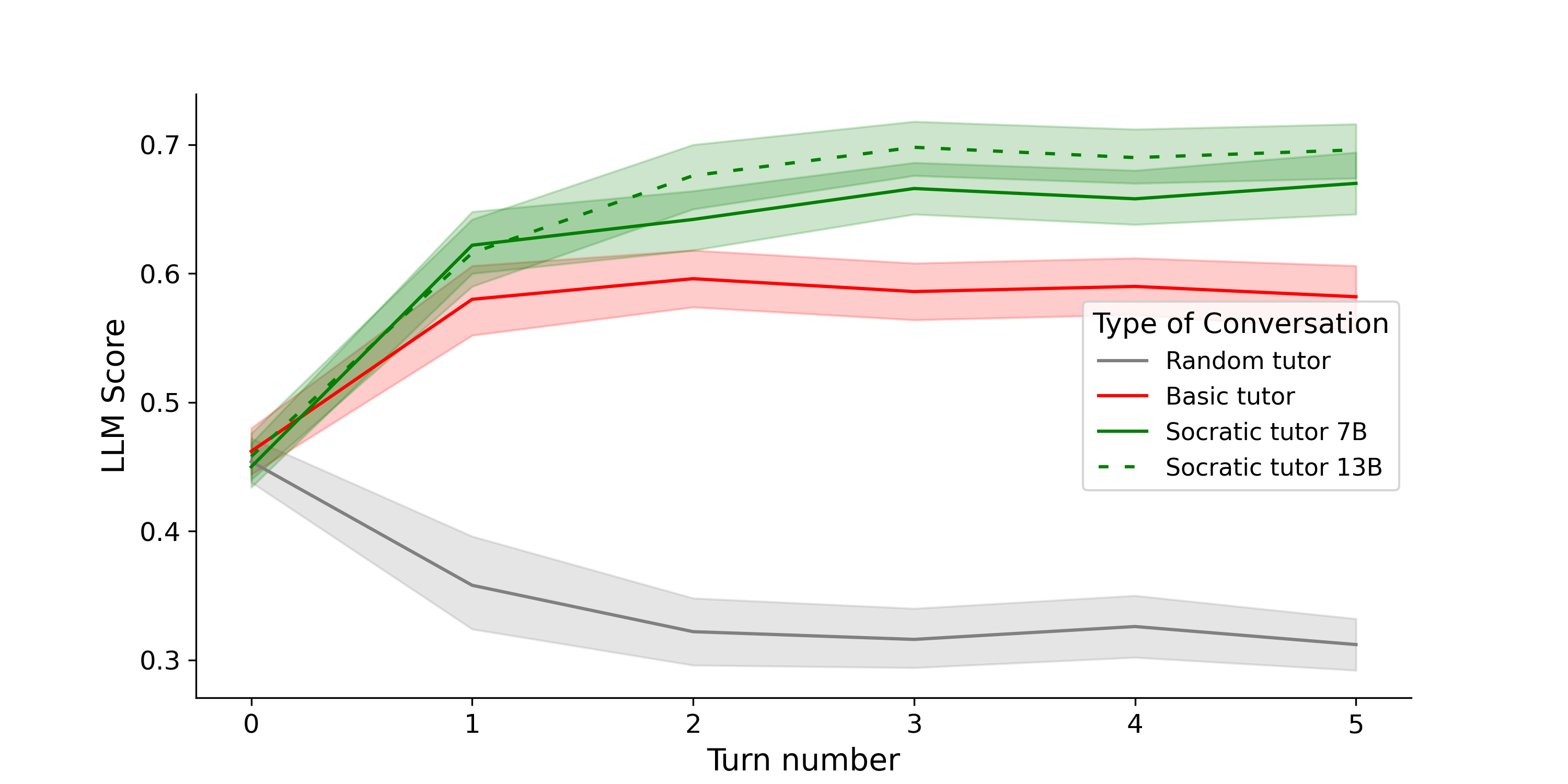}
    \caption{Critical thinking scores obtained by means of the LLM-score for the different types of tutors. The plots display the average values over 20 conversations of 5 turns for each of the 5 ToK questions. The differences between the Socratic tutor and the non-Socratic tutors are statistically significant (t-test, p-value<0.001). No statistically significant difference is observed in the performance of both Socratic tutors.}
    \label{fig:Performance2_llm}
\end{figure}

Figures~\ref{fig:meteor}, Figure~~\ref{fig:bertscore}, and Figure~\ref{fig:Performance2_llm} depict the performance (METEOR , BERT and LLM scores) of the different types of tutors across 5 turns for 20 conversations for each 5 of the ToK questions. Figures~\ref{fig:Performance2_bleu} and \ref{fig:Performance2_rouge}, in the supplementary material display the performance of the BLEU and ROUGE-L scores. As seen in the Figures, for all metrics and the Socratic and basic tutors, the scores increase with the number of turns. Furthermore, the \textit{Socratic tutor} significantly outperforms the other two tutors, demonstrating a more pronounced improvement as the conversation evolves. While the performance of the Socratic tutor implemented with the Llama2 13B-parameter model is superior to that of the Socratic tutor based on the Llama2 7B-parameter model, the differences in performance between these two tutors are not statistically significant.

\begin{table}[ht]
\caption{Performance of the different types of tutors, averaged over 20 conversations of 5 turns each for each of the 5 ToK questions. The differences between the Socratic tutors (highlighted in bold) and the non-Socratic tutors are statistically significant (t-test, p-value<0.001). The differences in performance between the two Socratic tutors are not statistically significant.} 
\centering
\begin{tabular}{p{1.5cm}p{0.7cm}p{0.7cm}p{0.7cm}p{0.7cm}p{0.7cm}}
\toprule
\textbf{\scriptsize{Tutor type}} &\textbf{\scriptsize{BLEU}}&\textbf{\scriptsize{ROUGE-L}}&\textbf{\scriptsize{METEOR}} &\textbf{\scriptsize{BERTScore}} &\textbf{\scriptsize{LLM Score}} \\ [1ex] \hline
Socratic Llama2 13B&\textbf{3.65}&\textbf{0.157}&\textbf{0.226}&\textbf{0.569}&\textbf{0.696}\\[0.9ex]
Socratic Llama2 7B &\textbf{3.42}&\textbf{0.162 }&\textbf{0.216}&\textbf{0.576}&\textbf{0.670}\\[0.6ex]
\hline
Basic Llama2 7B&0.494& 0.120 & 0.092&0.535&0.582\\[0.9ex] 
\hline
Random Llama2 7B & 0.210 & 0.091 &0.063&0.444&0.312\\

\bottomrule                
\end{tabular}
\label{tab:res}    
\end{table}

\section{Discussion} \label{sec:discussion}

From our study, we draw several implications. 

First, our results clearly demonstrate that the Socratic tutor outperformed both the basic and random tutors in assisting students with ToK questions. After only 5 turns of interaction, students engaging with the Socratic tutor showed a significant improvement in their critical thinking abilities compared to those interacting with the baseline chatbots. This superior performance indicates the effectiveness of employing Socratic questioning techniques in guiding students through the process of critical thinking.

Second, by developing educational chatbots based on small LLMs that can be run on personal computers, our methodology lowers the barriers of access to educational resources, democratizing education across different demographics and regions. Thus, our approach contributes to greater educational equity by enabling individuals to independently access and leverage advanced technological tools for learning, regardless of their socio-economic status or geographical location.

Third, as a preliminary evaluation, we assess the effectiveness of our method through dialogues with a simulated learner powered by an LLM. Our experimental methodology contributes to the growing use of LLMs to simulate human interactions in data generation tasks. Moving forward, we plan to conduct case studies involving actual students to further validate our method.

Fourth, despite the promising results obtained by leveraging fine and prompt-tuned pre-trained small LLMs for this task, we encountered challenges in prompt-tuning. Designing prompts that effectively elicit critical thinking and guide students through the reasoning process proved to be a complex task. Occasionally, our model would generate dialogues where the tutor both posed questions and provided answers, or indirectly guided the conversation by stating phrases such as \emph{``If I were a Socratic tutor, I would ask...'' or ``My response to the student is...''}. After extensive empirical work with different types of prompts, we discovered that removing the memory of the dialogue from the learner’s prompt substantially enhanced the coherence and logical flow of the conversations, leading to higher-quality interactions.

Fifth, surprisingly, the performance of the Socratic tutor based on the smallest LLM is at par with that of the Socratic tutor built from an LLM with almost twice the number of parameters. These results illustrate the power of even small LLMs to provide educational value even in contexts where resource constraints or computational limitations may be a concern. We leave to future work a more in-depth exploration of how to improve the performance of the Socratic tutor based on the Llama3 8B-parameter model or models of similar complexity. 

Sixth, our study's findings have significant implications for education, particularly in the context of online learning and intelligent tutoring systems. Integrating Socratic chatbots into educational platforms can provide students with personalized, interactive, and scaffolded learning experiences that promote the development of critical thinking skills. With our Socratic tutor, educators can augment traditional instruction methods and enhance students' ability to analyze complex issues, evaluate evidence, and construct well-reasoned arguments.

\section{Conclusion} \label{sec:conclusion}
Our study contributes to the development of AI-driven tutoring systems by introducing a fine- and prompt-tuned model specifically aimed at generating Socratic questions to enhance critical thinking skills. Our approach enables the creation of a Socratic tutor that surpasses the baselines in delivering effective tutoring conversations to foster critical thinking. The comparison highlights the potential of tailored AI approaches in educational settings.

Looking ahead, we propose several directions for future research. First, further research is needed to continue improving the effectiveness of our Socratic tutor in educational contexts, including the study of sentiment analysis, semantic understanding and reinforcement learning with human feedback to enhance the chatbot's ability to interpret and respond to students' queries more accurately. Second, our approach should be evaluated with human students to measure its effectiveness in real educational settings and gather qualitative feedback on its tutoring capabilities.
Third, there is a need to propose more refined evaluation methods that are able to 
adequately assess critical thinking in the context of automated tutoring conversations.  
Additionally, investigating the long-term impact of Socratic chatbot interventions on students' critical thinking skills and academic performance would provide valuable insights into the efficacy of this approach.
Furthermore, developing a learner model that adapts to individual student responses and learning progress could greatly enhance the personalization and effectiveness of the Socratic tutor. These steps will be crucial in advancing the field of AI-driven educational tools and in creating more effective, responsive and human-centered learning environments.

\section*{Acknowledgments}
This work has been funded by a nominal grant received at the ELLIS Unit Alicante Foundation from the Regional Government of Valencia in Spain (Convenio Singular signed with Generalitat Valenciana, Conselleria de Innovación, Industria, Comercio y Turismo, Dirección General de Innovación) and a grant by the Banc Sabadell Foundation.


\bibliography{mybibfile}

\clearpage
\section*{Supplementary material}
\subsection*{1. Ethical implications}
The experiments presented in this paper did not entail collecting any human data as the annotated dataset used for fine-tuning is publicly available from \cite{ang_socratic_2023} and the evaluation of the chatbots has been performed using LLM-based student agents. Hence, no IRB approval was needed. The main ethical aspects that we are implementing by design in the \texttt{Socratic} tutor include: 
\begin{enumerate}
    \item Preservation of privacy by leveraging open-source LLMs that are run locally on the students’ devices.
    \item  Non-manipulation by constraining the tutor to formulate Socratic questions, which are meant to foster critical thinking. 
    \item Factual veracity by focusing on formulating questions rather than providing answers.
    \item Inclusiveness by developing an open-source system that would be available to every student in their local devices.
\end{enumerate}

\subsection*{2. ToK questions and their summary}
Table~\ref{tab:ToK_eg_other} and Table~\ref{tab:ToK_eg_other_continued} present examples of Theory of Knowledge questions with a summary of the information that should be included in the answer.
\begin{table}[ht]  
\centering
\caption{ Examples of Theory of Knowledge question with a summary of the information that should be included in the answer.}  
\begin{tabular}{p{8
cm}}
\toprule
\textbf{For artists and natural scientists, which is more important: what can be explained or what cannot be explained?}\\[0.9ex] 
For artists and natural scientists, the question of what is more important, what can be explained or what cannot be expressed, is a complex and multifaceted one. Both groups rely heavily on the concept of explicability, or the ability to provide clear and comprehensible explanations for their work and findings. However, they approach this concept from different perspectives and have varying opinions on its significance.Artists often prioritize what cannot be explained over what can be explained. They recognize that much of their work is rooted in intuition, emotion, and personal experience, rather than objective facts or scientific laws. As such, they may view the limitations of language and expressibility as a liberating force, allowing them to tap into the richness of human experience and expression without being bound by the constraints of verbal explanation. In this sense, what cannot be explained becomes a source of creative freedom and inspiration.Natural scientists, on the other hand, tend to prioritize what can be explained over what cannot be explained. They are concerned with understanding the natural world through observation, experimentation, and mathematical modeling, and they recognize that their findings must be grounded in empirical evidence and logical reasoning. While they acknowledge the limits of language and expressibility, they see these limitations as a challenge to be overcome through rigorous scientific inquiry and communication. In this sense, what can be explained becomes a means of gaining knowledge and understanding about the world. Ultimately, both artists and natural scientists recognize that there are limits to their respective domains of explanation and expression, but they approach these limits in fundamentally different ways. While artists may view the unexplainable as a source of creative freedom, natural scientists see it as a challenge to be overcome through scientific inquiry. Nonetheless, both groups share a common goal of understanding and communicating their ideas effectively, even if they differ in their priorities and approaches.\\[2ex] 
\textbf{Does it matter if our acquisition of knowledge happens in "bubbles" where some information and voices are excluded?}\\[0.9ex] 
The question of whether it matters if our acquisition of knowledge happens in "bubbles" where some information and voices are excluded is a complex one. On one hand, it could be argued that there can be purely subjective knowledge, as individuals can gain insight and understanding through personal experiences and emotions that may not be shared with others. However, on the other hand, it is important to consider whether there can also be purely objective knowledge, which is independent of individual perspectives and biases.To answer this question, we must first define what we mean by "objective" and "subjective" knowledge. Objective knowledge refers to information that is verifiable and based on empirical evidence, while subjective knowledge is personal and internal, relying on an individual\'s personal experiences and beliefs.In terms of sharing another\'s perspective, it is important to recognize that understanding and empathy are crucial in building meaningful connections with others. To truly share someone else\'s perspective, one must be willing to listen actively, ask questions, and try to see things from their point of view. This requires a willingness to set aside one\'s own biases and assumptions, and to approach the situation with an open mind.However, it is also important to recognize that knowledge can be acquired through various means, including personal experiences, education, and cultural influences. These factors can shape an individual\'s perspective and understanding of the world, and it is possible for individuals to gain knowledge independently of one another.Ultimately, whether or not our acquisition of knowledge happens in "bubbles" does matter, as it can impact the diversity of perspectives and ideas that are represented. By seeking out a variety of sources of information and engaging in active listening and empathy, we can gain a more comprehensive understanding of the world around us. Additionally, recognizing the potential for both subjective and objective knowledge can help us to approach situations with a more nuanced understanding of the complexities involved. \\[2ex] 
\bottomrule
\end{tabular}
\label{tab:ToK_eg_other}    
\end{table}

\begin{table}[ht]  
\centering
\caption{ Examples of Theory of Knowledge question with a summary of the information that should be included in the answer, \textit{continued}.}  
\begin{tabular}{p{8cm}}
\toprule
\textbf{Do you agree that it is astonishing that so little knowledge can give us so much power?}\\[0.9ex] 
I agree that it is astonishing how little knowledge can give us so much power. This statement highlights the idea that there are various types or forms of power, and that knowledge does not necessarily have to be directly proportionate to power. In other words, having a limited amount of knowledge can still provide significant power and influence.The concept of "less is more" can apply here, where having less knowledge can sometimes lead to more power. For instance, a novice artist may create a masterpiece without the burden of extensive knowledge about art history or technical skills, allowing them to think outside the box and produce something unique and impactful. Similarly, a beginner entrepreneur may disrupt an industry by identifying untapped markets or unconventional business models without being weighed down by conventional wisdom. However, it is also possible to have power without knowledge. For example, someone with charisma and leadership skills can inspire and mobilize others without necessarily having extensive knowledge about their cause. A skilled orator can rally people around a vision or idea without needing to know every detail about the issue at hand.In conclusion, while there is an inverse relationship between knowledge and power in some cases, it is not always true that more knowledge leads to more power. Sometimes, less knowledge can result in more power, as it allows individuals to approach problems with fresh perspectives and unencumbered by conventional thinking.\\[2ex]
\textbf{Are visual representations always helpful in the communication of knowledge?}\\[0.9ex] 
Visual representations can be helpful in the communication of knowledge, but they are not always necessary or even the most effective means of conveying information. The relevance of truth to representation is crucial in determining whether a visual representation is useful. If the representation does not accurately reflect the truth, it may lead to confusion or misinterpretation, which can hinder knowledge communication.The distinction between practical and theoretical knowledge is also relevant when considering the use of visual representations. Practical knowledge is focused on solving problems and achieving specific goals, while theoretical knowledge is more concerned with understanding concepts and principles. Visual representations may be more effective in conveying practical knowledge, as they can provide a clear and concise visualization of complex information. However, theoretical knowledge may require more abstract and conceptual representations, such as text or diagrams, to fully convey the nuances of the subject matter. It is important to note that written language is not always a visual representation. While it may involve images or graphics, written language primarily consists of words and symbols that are used to convey meaning through the use of linguistic structures and conventions. Therefore, while visual representations can be helpful in certain contexts, they are not the only means of communicating knowledge, and written language can also be an effective tool for conveying information.\\[2ex] 
\bottomrule
\end{tabular} 
\label{tab:ToK_eg_other_continued}    
\end{table}

\subsection*{3. Other metrics: BLEU, ROUGE and METEOR}

Below, we summarize the advantages and limitations of the metrics: BLEU, ROUGE and METEOR.
\begin{itemize}
    \item \textbf{BLEU:} The Bilingual Evaluation Understudy or BLEU \cite{Papineni02bleu:a,lin-och-2004-orange} is a metric to evaluate the quality of machine-translated text by comparing it to one or more human-translated reference texts. Its highest possible value is 100. It measures the accuracy of the translation based on the overlap of words or phrases, where the larger the score, the better the match with the reference. However, by calculating average lengths across an entire corpus, BLEU potentially skews individual sentence scores. 
    \item \textbf{ROUGE:} The Recall-Oriented Understudy for Gisting Evaluation or ROUGE \cite{lin-2004-rouge} consists of a set of metrics designed to evaluate the output of automatic summarization and machine translation applications by comparing them to one or more human-generated reference summaries or translations. ROUGE scores range from 0 to 1, where higher values denote a larger resemblance between the machine-generated output and the human reference. Specifically, we compute the \textbf{ROUGE-L} score that utilizes statistics based on the \emph{longest common sub-sequence} approach. This method inherently accounts for similarities at the sentence level by identifying the longest sequence of words appearing in the same order within the automated output and the reference text.
    \item \textbf{METEOR:} The Metric for Evaluation of Translation with Explicit Ordering or METEOR \cite{banarjee2005} is an alternative evaluation metric specifically designed for machine translated texts. METEOR employs a weighted F-score that considers unigram mapping and incorporates a penalty function for incorrect word ordering. As the previous metric, it aims to align more closely with human judgment.
\end{itemize}

\subsection*{4. Performance by turn for other metrics}
\begin{figure}[ht] 
    \centering
    \includegraphics[width=1.1\linewidth]{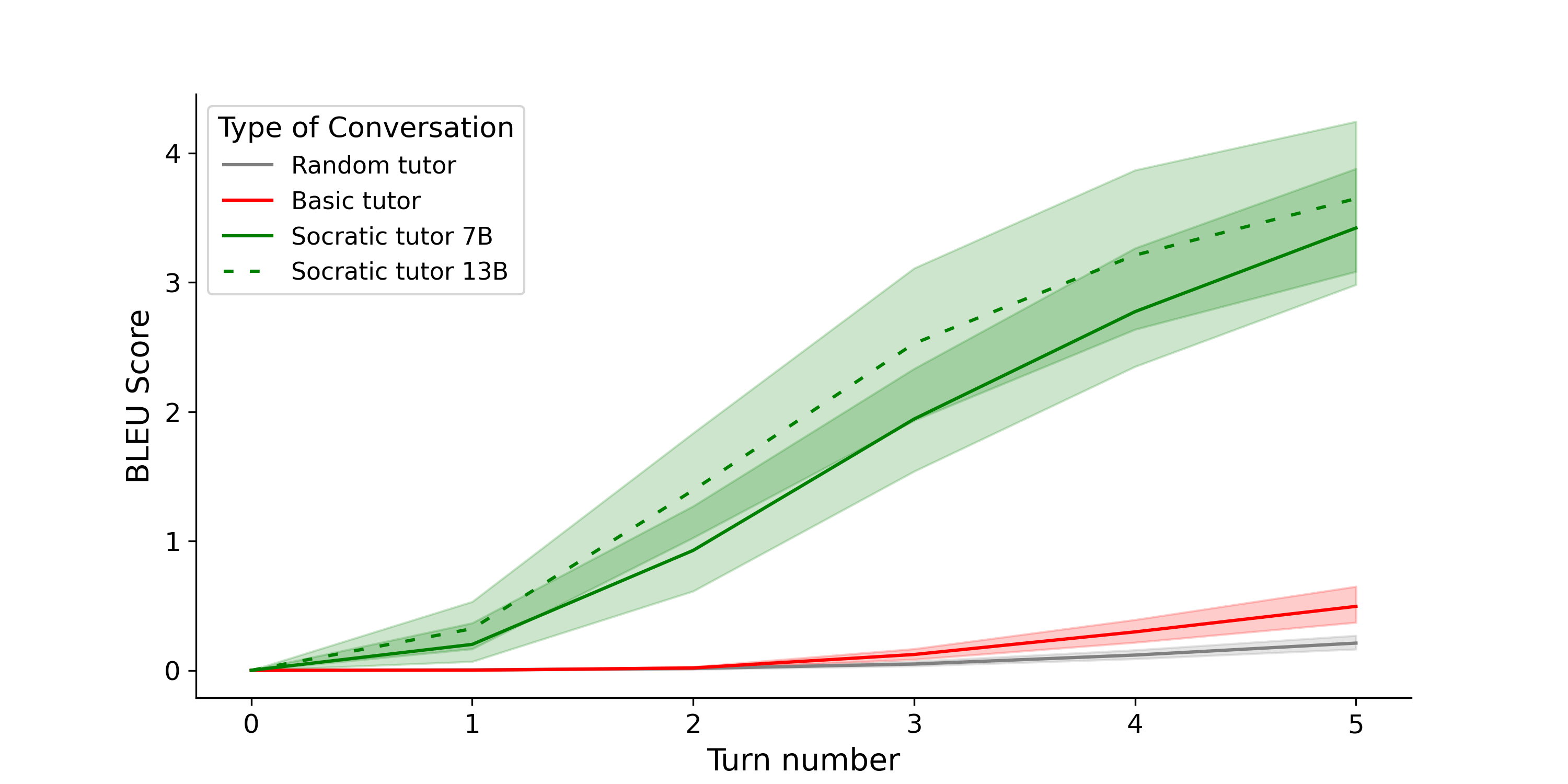}
    \caption{BLEU scores of the different types of tutors, averaged over 20 conversations of 5 turns each and for each of the 5 ToK questions. The differences between the Socratic tutors and the non-Socratic tutors are statistically significant (t-test, p-value<0.001). No statistically significant difference is observed in the performance of both Socratic tutors.}
    \label{fig:Performance2_bleu}
\end{figure}

\begin{figure}[ht] 
    \centering
    \includegraphics[width=1.1\linewidth]{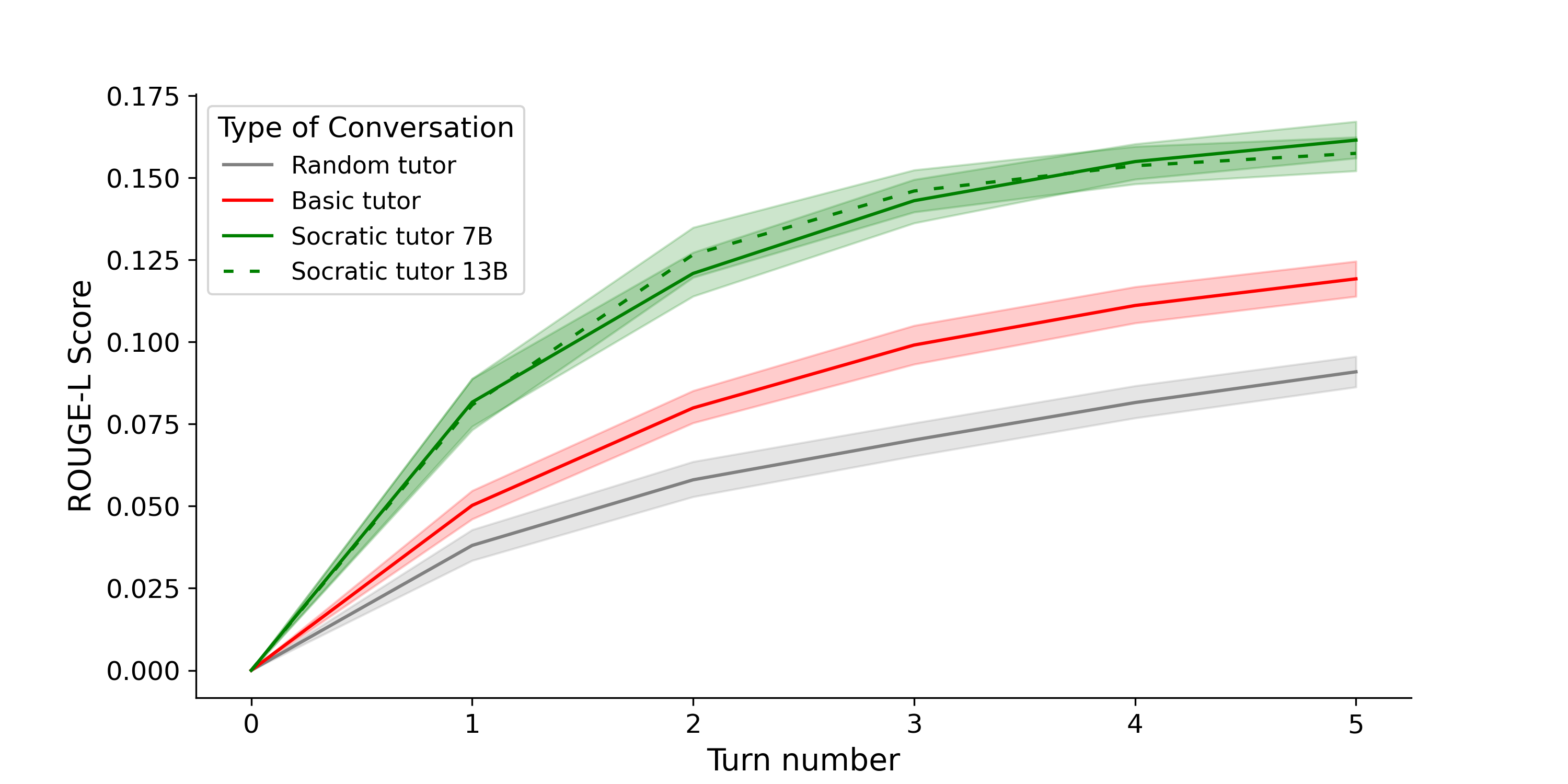}
    \caption{ROUGE-L scores of the different types of tutors, averaged over 20 conversations of 5 turns each and for each of the 5 ToK questions. The differences between the Socratic tutors and the non-Socratic tutors are statistically significant (t-test, p-value<0.001). No statistically significant difference is observed in the performance of both Socratic tutors.}
    \label{fig:Performance2_rouge}
\end{figure}

Figures~\ref{fig:Performance2_bleu} and~\ref{fig:Performance2_rouge} show the performance (BLEU and ROUGE-L) of the different types of tutors across 5 turns, averaged over 20 conversations for each 5 ToK questions.

\subsection*{5. Longer conversations}
\begin{figure}[ht] 
    \centering
    \includegraphics[width=1.1\linewidth]{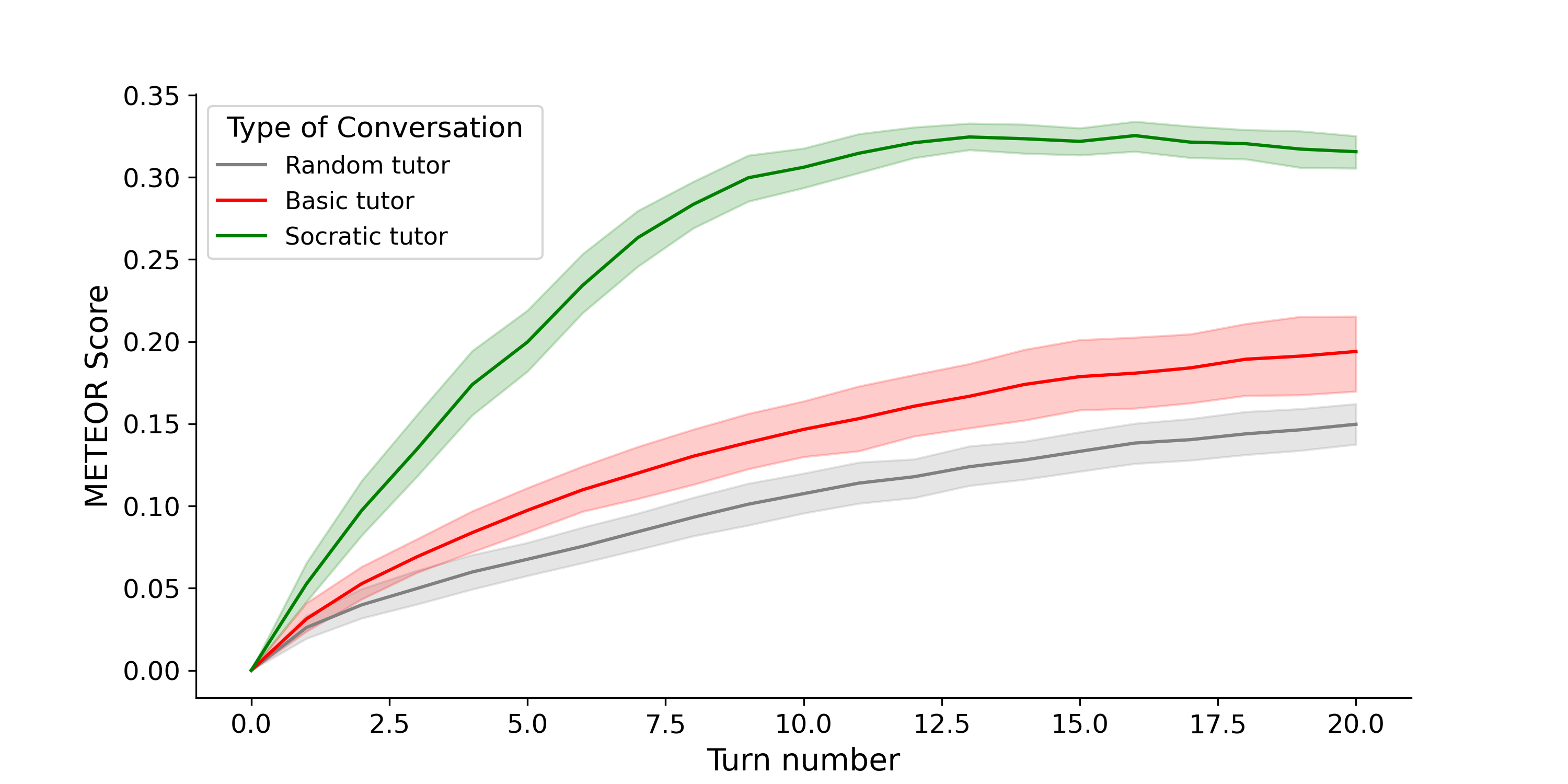}
    \caption{{METEOR scores of the different types of tutors, averaged over 20 conversations of 5 turns each and for each of the 5 ToK questions. The differences between the Socratic tutor (Llama2 7B parameters) and the non-Socratic tutors are statistically significant (t-test, p-value<0.001)}}
    \label{fig:Longer_conv}
\end{figure}

Figure~\ref{fig:Longer_conv} shows the performance (METEOR scores) of the different types of tutors across 20 turns, averaged over 20 conversations for each 5 ToK questions.

\subsection*{6. Prompt engineering}

 When we configured the prompts for the different LLMs, we observed significant sensitivity to the nuances of each prompt. Table~\ref{tab:prompt_engi} lists the main issues encountered and the corresponding prompts that resolved them.

\begin{table}[ht]  
\centering  
\caption{Main issues encountered and prompts that resolved them.} 
\begin{tabular}{p{2cm}p{4.5cm}}
\toprule
\multicolumn{1}{c}{\textbf{Issue}} & \multicolumn{1}{c}{\textbf{Prompt}}\\[0.5ex] \hline \\[0.1ex] 
Tutor responds by providing knowledge & "Use ONLY information of the dialogue history"\\[1ex] 
\hline \\[0.1ex] 
Generate long text & "ASK THE STUDENT ONE SHORT Socratic question" \\ 
                    & "ANSWER [...] IN ONE SENTENCE" \\ 
                     & "ONE SHORT SENTENCE" \\
                     & "generate a VERY SHORT" \\[1ex] 
\hline \\[0.1ex]                      
Repetition of the same question &  "DO NOT REPEAT YOURSELF"\\
                
\bottomrule
\end{tabular}
\label{tab:prompt_engi}    
\end{table}

\subsection*{7. Baseline Tutor}

 To further illustrate the value of the Socratic tutor, we ran additional experiments with a “baseline” tutor that hadn’t been neither fine- nor prompt-tuned. For 5 turns, we ask this baseline tutor: “Help me think about the question: <TOK question>” and assess its responses as we do with the other tutors. The difference in performance between the Socratic tutor and the baseline tutor is statistically significant for all the metrics (t-test, p-value<0.001).
\end{document}